\documentclass[a4paper,twoside,titlepage,12pt]{article}
\usepackage{amssymb}

\renewcommand{\baselinestretch}{1.5}

\newcommand{\be}{\begin{equation}}
\newcommand{\ee}{\end{equation}}
\newcommand{\bea}{\begin{eqnarray}}
\newcommand{\eea}{\end{eqnarray}}
\newcommand{\nn}{\nonumber}

\newcommand{\spa}{\ \ \ }
\newcommand{\eqref}[1]{(\ref{#1})}
\newcommand{\tr}{\mathop{{\rm Tr}}}
\newcommand{\bo}{\mathbf}
\newcommand{\conus}{\cos (\omega t)}
\newcommand{\sinus}{\sin (\omega t)}

\begin{document}
\begin{titlepage}
      
  \rightline{NBI-HE-00-nn}
  \rightline{hep-th/0009029}
  \rightline{August, 2000}
  
  \vskip 2cm
  \centerline{\Large \bf Stability of the}
  \vskip 0.2cm
  \centerline{\Large \bf Rotating Ellipsoidal D0-brane System} 
  
  \vskip 0.4cm

    \vskip 0.3cm
  \centerline{{\bf Konstantin G. Savvidy}\footnote{e-mail: savvidis@nbi.dk} }
  
\centerline{\sl The Niels Bohr Institute, Blegdamsvej 17, DK-2100 Copenhagen \O, Denmark}
  

 \centerline{{\bf George K. Savvidy}\footnote{e-mail: savvidy@mail.demokritos.gr}  }

 \centerline{\sl National Research Center Demokritos,
 Agia Paraskevi, GR-15310 Athens, Hellenic Republic }

\vskip 2cm
  \centerline{\bf Abstract}
  In this note we prove the complete stability of the classical 
fluctuation modes
of the rotating ellipsoidal membrane. The analysis is 
carried out in the full $SU(N)$ setting, with the conclusion that the
fluctuation matrix has only positive eigenvalues. This proves that
the solution will remain close to the original one for all time,
under arbitrary infinitesimal perturbations of the gauge fields.
  \vskip 0.4cm
\noindent

\end{titlepage} 

\section{Introduction}
\label{sec:intro}

The purpose of this note is to make a complete analysis of the
gauge field fluctuations in the neighbourhood of the rotating ellipsoidal
membrane solution of \cite{Harmark:2000na}.  
We extend the previous treatment in \cite{Harmark:2000na} whereby only
perturbations that do not modify the original $SU(2)$ ansatz to the case when
perturbations are in the full $SU(N)$ algebra.  
The results
indicate that in the case of $SU(2)$ most of the modes display the
enhanced symmetry of the original solution, {\it i.e.} after the imposition of 
the constraint most of the additional degrees of freedom are zero-modes.
All the other modes, for the totality of all possible gauge field
perturbations in $SU(N)$, are completely stable and execute harmonic
oscillations around the original trajectory.

The effective action of $N$ D0-branes for weak and slowly
varying fields is the non-abelian SU(N) Yang-Mills action plus the
Chern-Simons action (for the bosonic part).  For weak fields the
action is gotten by dimensionally reducing the action of 9+1
dimensional $U(N)$ Super Yang-Mills theory to 0+1 dimensions
\cite{Witten:1996im}.  Up to a constant term it is
\begin{equation}
\label{BIaction}
S_{} = - T_0 (2\pi l_s^2)^2 \int dt \, 
 \tr \left(\frac{1}{4} F_{\mu \nu} F^{\mu \nu} \right)~~,
\end{equation}
where $F_{\mu \nu}$ is the non-abelian $U(N)$ field strength
in the adjoint representation and \( T_0 = ( g_s l_s )^{-1} \)
is the D0-brane mass.
To write this action in terms of coordinate matrices \( \bo{X}^i \),
one has to use the dictionary 
\begin{equation}
A_i = \frac{1}{2\pi l_s^2} \bo{X}^i  ,\spa
F_{0i} = \frac{1}{2\pi l_s^2} \dot{\bo{X}}^i  ,\spa
F_{ij} = \frac{-i}{(2\pi l_s^2)^2} \, [\bo{X}^i,\bo{X}^j]
\label{eq:dict}
\end{equation}
with \( i,j = 1,2,...,9 \), giving 
\begin{equation}
\label{ham}
S_{} = T_0 \int dt \, \tr \left(
\frac{1}{2} \dot{\bo{X}}^i \dot{\bo{X}}^i
+ \frac{1}{4} \frac{1}{(2\pi l_s^2)^2 } [\bo{X}^i,\bo{X}^j][\bo{X}^i,\bo{X}^j] \right)
\end{equation}
To derive this it is necessary to gauge the $A_0$ potential  away,
which is possible for a non-compact time. 
The equations of motion 
\be
\label{eq:eom}
\ddot{\bo{X}}^i = - \left[ \bo{X}^j \, ,\left[ \bo{X}^j,\bo{X}^i \right] \, \right]
\ee
should be taken together with the Gauss constraint, 
\begin{equation}
\label{eq:gauss}
\left[\dot{\bo{X}}^i,\bo{X}^i \right] = 0
\end{equation}
which is preserved by (\ref{eq:eom}.

Introduce the $N \times N$ matrices
$\mathbf{L}_1,\mathbf{L}_2,\mathbf{L}_3$ as the generators of the $N$
dimensional irreducible representation of $SU(2)$, with algebra
\begin{equation}
\label{SU2alg}
\left[\bo{L}_i,\bo{L}_j\right] = i \, \epsilon_{ijk} \bo{L}_k~~.
\end{equation}

The rotating spherical membrane solution was constructed using this
subalgebra of the $SU(N)$. It is also the only finite dimensional
subalgebra of the group of diffeomorphisms of $S^2$, the Sdiff$(S^2)$.
That is why the $SU(2)$ ansatz is in some sense unique:
it is the only type of solution that carries  over to the 
supermembrane without modification
\begin{eqnarray}
\label{ansatz6}
\bo{X}^i(t) &=& \bo{L}_i \, R \, \cos(\omega t)\\
\tilde{\bo{X}}^i(t) = \bo{X}^{i+3}(t) &=& \bo{L}_i \, R \, \sin(\omega t)
\end{eqnarray}
In what follows we will set $R=1$, with $\omega$ being determined through the
equations of motion, and equal to $\omega^2 = 2R^2 =2$.

In order to treat the perturbations of the system within the full
$SU(N)$ we will need a convenient basis, provided \cite{hoppe,deWit:1989ct} by 
the spherical operators $\bo{Y}^l_m$, 
\bea
\left[\bo{L}_z, \bo{Y}^l_m\right] &=& m \, \bo{\bo{Y}}^l_m 
\qquad\qquad\qquad \textrm{for ~~}l=1,\ldots,N-1\nn \\
\left[\bo{L}_{\pm} ,{\bo{Y}}^l_m \right] &=& \sqrt{ (l \mp m)(l \pm m +1) } \, {\bo{Y}}^l_{m\pm1}
\label{eq:Ydef}
\eea
It is immediately clear that these are not hermitian, neither they are
anti-hermitian.
For example, at $l=1$ these  should coincide with the $\bo{L}_i$'s:
\bea
\bo{Y}^1_0 = \bo{L}_z       ~~~,~~~~~~~~~
\bo{Y}^1_{\pm 1} = \mp {1\over\sqrt{2}} \, \bo{L}_{\pm} ~.
\eea
For $l=2$ these are the five linear independent symmetric traceless
products of pairs of $\bo{L}$'s, and with correct normalization read:
\bea
\bo{Y}_0^2 &=& 
\sqrt{ {2 \over 3} }\, ( \bo{L}_x \bo{L}_x + \bo{L}_y \bo{L}_y - \bo{L}_z \bo{L}_z ) \nn\\
\bo{Y}_{\pm 1}^2 &=& \pm \, \{ \bo{L}_{\pm}, \bo{L}_z \}   \nn\\
\bo{Y}_{\pm 2}^2 &=& ( -\bo{L}_x \bo{L}_x + \bo{L}_y \bo{L}_y \mp 2 i \, \{\bo{L}_x, \bo{L}_y \} )
\eea
The general explicit construction of the $\bo{Y}^l_m$ is due to 
Schwinger, and was used to show the correspondence at $N \mapsto \infty$
between the relativistic membrane and $SU(N)$ YM in \cite{hoppe,deWit:1988ig,deWit:1989ct}.
We will not use the explicit form of these matrices, as the defing
relations \eqref{eq:Ydef} is all that is needed.
The properties under Hermitian conjugation can be summed up as
\be
\bo{Y}_{l,m}^{\dagger} = (-1)^m \, \bo{Y}_{l,-m}
\ee

\section{Internal Perturbations}
\label{sec:int_pert}
In this section perturbations will be considered that are parallel 
in space to one of the directions of the system. It is a generalization of 
our previous treatment, in \cite{Harmark:2000na}, where perturbations 
of the same structure as the ansatz were considered. The case of completely
transverse perturbations is considered in the next section.
The details are rather technical, but completely straightforward.

Let us decompose the fluctuation fields in the basis defined by 
$\bo{Y}^l_m$, where $l$ runs from $1$ to $N-1$ 
\be
\label{eq:eta-def}
\delta \bo{X}^i = \sum_{m=-l}^l \bo{Y}^l_m \, \xi_m^i ~~,\qquad \delta
\tilde{\bo{X}}^i = \sum_{m=-l}^l \bo{Y}^l_m \, \eta_m^i \qquad
i=1,2,3.  
\ee 
The total number of modes is then $\sum_{l=1}^{N-1}
(2l+1) = N^2-1$ as it should for $SU(N)$. We do not introduce an $l$
index on the $\eta,\xi$ because it will be shown below, and also
suggested in \cite{Axenides:2000mn}, that the fluctuations with
different $l$ do not couple at linear order.  The behaviour of linear
perturbations is sufficient to establish the correct phase portrait of
the dynamical system in the neighbourhood of a periodic trajectory.

Even though the basis is not Hermitian, 
the gauge field should still be real, and so we should impose 
\be
\label{eq:real}
\xi_m^{*} = (-1)^m  \xi_{-m} ~~~\textrm{and}~~~~~~ \eta_m^{*} = (-1)^m  \eta_{-m} ~~\textrm{for all}~~m=-l,\ldots,l~~.
\ee
The variational equations of motion are
\bea
\label{eq:var-eom}
- \delta \ddot{\bo{X}}^i &=& 
\left[ \delta \bo{X}^j \, ,\left[ \bo{X}^j,\bo{X}^i \right] \, \right] +
\left[ \bo{X}^j \, ,\left[ \delta \bo{X}^j,\bo{X}^i \right] \, \right] +
\left[ \bo{X}^j \, ,\left[ \bo{X}^j, \delta \bo{X}^i \right] \, \right] + \nn\\
&&\left[ \delta \tilde{\bo{X}}^j \, ,\left[ \tilde{\bo{X}}^j,\bo{X}^i \right] \, \right] +
\left[ \tilde{\bo{X}}^j \, ,\left[ \delta \tilde{\bo{X}}^j,\bo{X}^i \right] \, \right] +
\left[ \tilde{\bo{X}}^j \, ,\left[ \tilde{\bo{X}}^j, \delta \bo{X}^i \right] \, \right] ~~.
\eea
The linearized constraint equation looks like
\be
\label{eq:var-constr}
\sum_{i,m} \left[\delta \dot{\bo{X}^i} , \bo{X}^i\right] + \left[ \dot{\bo{X}^i} , \delta \bo{X}^i\right] + 
\left[\delta \dot{\tilde{\bo{X}}^i} , \tilde{\bo{X}}^i\right] + 
\left[ \dot{\tilde{\bo{X}}^i} , \delta \tilde{\bo{X}}^i\right]  = 0
\ee
Using the commutation relations (\ref{eq:Ydef}) we get for the constraint
\be
\sum_i \bo{L}^i_{nm} 
\left( \conus \dot{\xi}_m^i + \omega \sinus {\xi}_m^i + 
\sinus \dot{\eta}_m^i - \omega \conus \eta_m^i \right) =0~~,
\ee
where $\bo{L}^i_{nm}$ are now the $SU(2)$ generators in the 
$(2l+1)\times(2l+1)$ representation. 
In the co-moving  coordinates 
\be
\label{eq:comov}
u_m^i= \conus \xi_m^i + \sinus \eta_m^i~~~
\textrm{and}~~~~ v_m^i = -\sinus \xi_m^i +\conus \eta_m^i
\ee
the constraint looks simpler,
\be
\sum_{i,m} \bo{L}^i_{nm} \left ( \dot{u}_m^i -2 \omega v_m^i \right ) =0~.
\label{constr}
\ee
The variational equation of motion (\ref{eq:var-eom}) after substituting
the fields (\ref{eq:eta-def}) is
\bea
- \sum_m \bo{Y}^l_m \ddot{\xi}_m^i = & \sum_{j,k,m}
&{}- \conus \, \conus  i \, \epsilon_{ijk} \left[\bo{Y}^l_m , \bo{L}_k\right] \xi_m^j      \nn\\
&&{}+ \conus \, \conus \left[\bo{L}_j\left[\bo{Y}^l_m,\bo{L}_i\right]\right] \xi_m^j  \nn\\
&&{}+ \conus \,  \conus \, l(l+1) \,\bo{Y}^l_m \xi_m^i                                          \nn\\
&&{}- \conus \,  \sinus  i \, \epsilon_{ijk} \left[\bo{Y}^l_m , \bo{L}_k\right] \eta_m^j    \nn\\
&&{}+ \conus \,  \sinus \left[\bo{L}_j\left[\bo{Y}^l_m,\bo{L}_i\right]\right] \eta_m^j \nn\\
&&{}+ \sinus \,  \sinus \, l(l+1) \, \bo{Y}^l_m \eta_m^i
\eea
and in component form,
\be
-\ddot{\xi}_n^i = 
\conus \, 
\left( i \, \epsilon_{ijk} \bo{L}^k_{nm} - \bo{L}^j_{n n^{\prime} }\bo{L}^i_{n^{\prime} m} \right) \,
\left( \conus \, \xi_m^j +  \sinus \, \eta_m^j \right)  + l(l+1) \, \xi_n^i~.
\ee
The decoupling of the modes with different $l$ is seen to be a direct
consequence of (\ref{eq:Ydef}), and more fundamentally, of the pure $SU(2)$
structure of the original background solution.

The above can be conveniently rewritten as
\be
\ddot{\xi}_n^i + l(l+1) \,  \xi_n^i = 
 \conus \, \left( \bo{L}^j_{n n^{\prime} }\bo{L}^i_{n^{\prime} m} +
 i \, \epsilon_{jik} \bo{L}^k_{nm} \right) \,
\left( \conus \, \xi_m^j +\sinus \, \eta_m^j \right) ~.               
\ee
The equation for $\eta$ is gotten by exchanging cosines for sines and $\xi$ for $\eta$
\be
\ddot{\eta}_n^i  + l(l+1) \, \eta_n^i = 
  \sinus \, \left( \bo{L}^j_{n n^{\prime} }\bo{L}^i_{n^{\prime}
  m} + i \, \epsilon_{jik} \bo{L}^k_{nm} \right) \, 
  \left( \conus \, \xi_m^j + \sinus \, \eta_m^j \right) ~.
\ee
In the co-moving  coordinates (\ref{eq:comov})
the time dependency drops out, and the equation
becomes a linear system with constant coefficients:
\bea
\ddot{u}_n^i + \left( l(l+1) -2 \right) u_n^i -2\omega \, \dot{v}_n^i &=& 
\left( \bo{L}^j_{n n^{\prime} }\bo{L}^i_{n^{\prime} m} +
 i \, \epsilon_{jik} \bo{L}^k_{nm} \right) \, u_m^j ~, \label{eq:fnl1}\\
\ddot{v}_n^i + \left( l(l+1) -2 \right) \, v_n^i + 2\omega \, \dot{u}_n^i  &=& 0 ~. 
\label{eq:fnl2}
\eea
Thus we shall analyze the system of equations (\ref{constr}) 
(\ref{eq:fnl1}),  (\ref{eq:fnl2}).  
In order to display explicitly the constant matrix structure of the equation,
one has to write the $rhs$ of \eqref{eq:fnl1} as a matrix acting on a $3(2l+1)$ component vector
\be
        \left(\begin{array} {ccc}
        \bo{L}^1 \bo{L}^1 & \bo{L}^2 \bo{L}^1-i\bo{L}^3 & \bo{L}^3 \bo{L}^1 +i \bo{L}^2 \\
        \bo{L}^1 \bo{L}^2+i \bo{L}^3 & \bo{L}^2 \bo{L}^2 & \bo{L}^3 \bo{L}^2 -i\bo{L}^1\\
        \bo{L}^1 \bo{L}^3 -i\bo{L}^2 & \bo{L}^2 \bo{L}^3+i \bo{L}^1 & \bo{L}^3 \bo{L}^3 
        \end{array}\right)  
        \left(\begin{array}{c}
        u^1 \\ u^2 \\ u^3
        \end{array}\right)~.
 \ee
The eigenvalues $\Lambda$ of this size  $3(2l+1)$ block matrix,
where each block is of size $(2l+1)$, are given in the table, together
with their multiplicity
%
\be
\label{eq:Lambda}
\begin{array}{|c|c|}
\Lambda         & \textrm{multiplicity}     \\
\hline
-2l-2           &    2l-1                   \\
  l(l+1) -2      &    2l+1                   \\
  2 l             &    2l+3                   \\
\hline
\end{array}
\ee
One can check that the trace of the  matrix matches with the \emph{weighted} 
sum of the eigenvalues given above for arbitrary values of $l$. The proof
will be published elsewhere.

Since the second equation \eqref{eq:fnl2} is completely diagonal with respect to $i,l,m$, 
we can now solve the complete system. Choose  a fixed frequency ansatz
\be
u^i_n(t) = e^{ i \Omega t} \, u^i_n ~~,~~~v^i_n(t) = e^{ i \Omega t} \, v^i_n ~~.
\ee
The second equation \eqref{eq:fnl2} can be solved as
\be
v^i_n = \frac{ -2\sqrt{2} \, i \, \Omega^2} { l(l+1) -2 - \Omega^2} \,  u^i_n ~~,
\ee
and substituted back into the first equation \eqref{eq:fnl1},
\be
- \Omega^2 u^i_n +  (l(l+1) -2) u^i_n - \frac{ 8 \, \Omega^2} { l(l+1) - 2 - \Omega^2} \, u^i_n =  
\left(
\bo{L}^j_{n n^{\prime} }\bo{L}^i_{n^{\prime} m} +
 i \, \epsilon_{jik} \bo{L}^k_{nm} 
\right) \,  u^j_m ~.
\ee
In the basis in which the matrix on the $rhs$ is diagonalized,
it can be replaced with its respective eigenvalue $\Lambda$,
resulting in an algebraic equation for the $\Omega$
\be
\left (l\left (l+1\right )-2-\Omega^2 \right )^{2}-8\,\Omega^2 =\Lambda\,
\left (l\left (l+1\right )-2-\Omega^2 \right )~.
\ee
Finally, this quadratic equation can be solved, 
\be
\Omega_{1,2}^2  = -{1\over 2}\,\Lambda+ l(l+1)+2 
\pm {1\over 2}\,\sqrt {{\Lambda}^{2}-16\,\Lambda+32\, l(l+1) }~.
\ee
For the  values of $\Lambda$, taken from table (\ref{eq:Lambda}),
the modes are:
\be
\label{eq:modes}
\begin{array}{|c|c|c|c|}
\Lambda & \Omega^{2}_{1} & \Omega^{2}_{2} & \textrm{multiplicity} \\
\hline
 l(l+1) -2 &   0              & l^2 + \, l +6    & 2l+1             \\
 2l        &   l^2 -3l +2     & l^2 + 3l +2    & 2l+3           \\
 -(2l+2)   &   l^2 -\, l~     & l^2 + 5l +6    & 2l-1             \\
\hline
\end{array}
\ee
Note that the number of zero modes changes from $9$ for the case $l=1$, and $12$ for $l=2$,
to $2l+1$ for arbitrary $l>2$. 
This is connected with the fact that the original solution
is based on an $l=1$ ansatz, and so the symmetries of the equations are manifested 
as zero-modes under $l=1$ perturbations.
Thus, for $SU(N)$, the total number of zero modes is
the sum for all $l$ up to $N-1$
\be
9 + 12 + \sum_{l=3}^{N-1} (2l+1) = N^2 + 12
\ee
where $l$ runs up to $N-1$. 

Note that because the frequencies are real, we can indeed satisfy 
the gauge field reality conditions \eqref{eq:real} by choosing initial vectors
that satisfy reality. 
In addition the constraint conditions should be imposed, with the result 
that the $2l+1$ modes with frequency $l^2 + \, l +6$ are projected out at each $l$.

\section{Transverse perturbations}
\label{sec:trans}

In addition to the already considered perturbations there are also those that are
completely transverse to the system. That is the directions 789, if we had oriented the
original system along 123456. The analysis is considerably simpler that the previous case.
The perturbations
\be
\delta\bo{X}^k = \sum_m  \bo{Y}^l_{m} \,  \zeta_m^k ~~~~~\textrm{for}~~~k=7,8,9
\ee
satisfy the simple harmonic equation
\be
\ddot{\zeta}_m^k + l(l+1) \, {\zeta}_m^k = 0 ~~.
\ee
This clearly has only positive frequencies and is therefore stable. For $l=1$
all the $9$ modes have the same frequency as the original solution, corresponding
to infinitesimal global rotations of the system into the 789 hyperplane. 
The counting goes as follows, there are $9\times 2 =18$ first order degrees of
freedom here, which coincides with the dimensionality of the grassmanian manifold
of embeddings of a 6-hyperplane into $\mathbb{R}^9$, i.e. $ \frac{SO(9)}{ SO(6) \times SO(3) } $.

\section{Conclusion}
\label{sec:ack}
From the results of the two previous sections it follows that,
zero-modes notwithstanding, all the frequencies in the system are
positive, and arbitrary small perturbation will remain bounded for all
times.  We have learned also of the paper \cite{Axenides:2000mn}
where the same problem is considered in the membrane language.
However the authors of \cite{Axenides:2000mn} in their approach to the same problem 
arrive at the Mathiew equation  instead of the equations 
(\ref{eq:fnl1}),  (\ref{eq:fnl2}), (\ref{constr}) 
and therefore to the opposite conclusion, namely 
that there exist solutions to the linearized perturbation equations
which grow exponentially. We hope that the present work will
contribute to the clarification of the question.

\section{Appendix}

Here we shall present  stability analysis of the $l=1$  system  
made in \cite{} and shall 
compare it with the general consideration in the main text. In addition we shall find new 
solutions. The Hamiltonian of the system is 
\cite{Harmark:2000na,Savvidy:2000pd} 
\be
H= {1 \over 2} \sum^{6}_{i=1} \dot{r}^{2}_{i} + {1 \over 2} [(r^{2}_{1} +r^{2}_{2})(r^{2}_{3} +r^{2}_{4})
+ (r^{2}_{3} +r^{2}_{4})(r^{2}_{5} +r^{2}_{6}) + (r^{2}_{5} +r^{2}_{6})(r^{2}_{1} +r^{2}_{2})]
\ee
where $$\bo{X}^{i+1} =  {\bo L}_{1 + i/2}~ r_{i+1},~~\bo{X}^{i+2} =
~{\bo L}_{1+i/2}~ r_{i+2},~~i=0,2,4.$$ 
It is convenient to introduce new coordinates 
$$
\begin{array}{ll}
r_{1} = \rho_1 ~\cos \phi_1,&r_{2} = \rho_{1} ~\sin \phi_{1},\\
r_{3} = \rho_2 ~\cos \phi_2,&r_{4} = \rho_{2} ~\sin \phi_{2},\\
r_{5} = \rho_3 ~\cos \phi_3,&r_{6} = \rho_{3} ~\sin \phi_{3},
\end{array}
$$
so that the 
Hamiltonian take the form:
\be
H= {1 \over 2} \sum^{3}_{i=1} \left[\dot{\rho}^{2}_{i} + \rho^{2}_{i} \dot{\phi}^{2} \right]
 + {1 \over 2} [\rho^{2}_{1}\rho^{2}_{2} + \rho^{2}_{2}\rho^{2}_{3} + 
 \rho^{2}_{3}\rho^{2}_{1}].
\ee
The conservation integrals are:
$$
\rho^{2}_{1}\dot{\phi}_{1} = M_{1},~~~\rho^{2}_{2}\dot{\phi}_{2} = M_{2},~~~\rho^{2}_{3}\dot{\phi}_{3} = M_{3},
$$
and the effective Hamiltonian take the form: 
\be
H = {1 \over 2} \sum^{3}_{i=1} 
\left[ \dot{\rho}^{2}_{i} + {M^{2}_{i} \over \rho^{2}_{i}} \right] +
{1 \over 2} [\rho^{2}_{1}\rho^{2}_{2} + \rho^{2}_{2}\rho^{2}_{3} + \rho^{2}_{3}\rho^{2}_{1}].
\ee
The effective potential is equal to 
$$
U = {1 \over 2} [ {M^{2}_{1} \over \rho^{2}_{1}} + {M^{2}_{2} \over \rho^{2}_{2}} + {M^{2}_{3} \over \rho^{2}_{3}}] +
{1 \over 2} [\rho^{2}_{1}\rho^{2}_{2} + \rho^{2}_{2}\rho^{2}_{3} + \rho^{2}_{3}\rho^{2}_{1}]~.
$$
The equations of motion are :
\bea
\ddot{\rho_{1}} = -\rho_{1} (\rho^{2}_{2} +\rho^{2}_{3} ) 
+{ M^{2}_{1} \over \rho^{3}_{1} }  \\
\ddot{\rho_{2}} = - \rho_{2} (\rho^{2}_{1} +\rho^{2}_{3} ) 
+ { M^{2}_{2} \over \rho^{3}_{2} } \\
\ddot{\rho_{3}} = -\rho_{3} (\rho^{2}_{1} +\rho^{2}_{2} ) 
+{ M^{2}_{3} \over \rho^{3}_{3} }
\eea
and the previous solution \cite{} is $\rho_{i} = R_{i} = Const, i=1,2,3$ and $\dot{\phi}^{2}_{i}
=\omega^{2}_{i} = M^{2}_{i}/R^{4}_{i}= R^{2}_{i+1} + R^{2}_{i+2}$.
Let us now consider the special case when all coordinate are equal to each other
$\rho_{1} =\rho_{2} =\rho_{3} =\rho(t) $ and can depend on time,  then 
$$
H= {3 \over 2}[ \dot{\rho}^{2} + {M^2 \over \rho^{2}}  + \rho^4 ]
$$
and corresponding equation can be integrated. The new solution is elliptic function
$\rho = \rho(t)$ 
$$
t = \int^{\rho(t)}_{\rho_{min}}{d\rho \over \sqrt{2E/3 - \rho^4 - M^2/\rho^2}}
$$
with period 
$$
T = 2 \int^{\rho_{max}}_{\rho_{min}}{d\rho \over \sqrt{2E/3 - \rho^4 - M^2/\rho^2}},
$$
where $\rho_{min}, \rho_{max}$ are the solutions of the equation 
$2E/3 - \rho^4 - M^2/\rho^2 = 0.$ 

Let us now turn to a stability analysis of the solution 
$\rho_{1} =\rho_{2} =\rho_{3} =R =Const$ considered in the main text. 
The equations of variation are $(M^{2}_{i} = 2R^6)$:
\bea
\delta\ddot{\rho_{1}} = -2 R^2 (\delta \rho_{1} +  \delta \rho_{2} +\delta \rho_{3} ) 
- {3 M^{2}_{1} \over R^4 } \delta \rho_{1} = 
- R^2 (8 \delta \rho_{1} + 2 \delta \rho_{2} + 2\delta \rho_{3} )  \\
\delta\ddot{\rho_{2}} = - 2 R^2 (\delta \rho_{1} +  \delta \rho_{2} +\delta \rho_{3} ) 
- {3 M^{2}_{2} \over R^4} \delta \rho_{2} = 
- R^2 (2 \delta \rho_{1} + 8 \delta \rho_{2} + 2\delta \rho_{3} ) \\
\delta\ddot{\rho_{3}} = -2 R^2 (\delta \rho_{1} +  \delta \rho_{2} +\delta \rho_{3} ) 
- {3 M^{2}_{3} \over R^4 } \delta \rho_{3} = 
- R^2 (2 \delta \rho_{1} + 2 \delta \rho_{2} + 8\delta \rho_{3} ) 
\eea
with three stable modes $12,6,6$. They coincide with the ones in (\ref{eq:modes}). For the 
more general case when $R_{1}\neq R_{2} \neq R_{3}$ we have:
\bea
\delta\ddot{\rho_{1}} = -4 (R^{2}_{2}  + R^{2}_{3})\delta \rho_{1} 
- 2R_{1}R_{2}  \delta \rho_{2} - 2R_{1}R_{3}  \delta \rho_{3} \\
\delta\ddot{\rho_{2}} =  - 2R_{2}R_{1}    \delta \rho_{1} 
-4 (R^{2}_{1}  + R^{2}_{3})\delta \rho_{2} - 2R_{2}R_{3} \delta \rho_{3} \\
\delta\ddot{\rho_{3}} = - 2R_{3}R_{1}\delta \rho_{1} - 2R_{3}R_{2} \delta \rho_{2}
-4 (R^{2}_{1}  + R^{2}_{2})\delta \rho_{3},
\eea
which also has only positive modes \cite{Savvidy:2000pd}.

\addcontentsline{toc}{section}{References}
\renewcommand{\baselinestretch}{1}

\bibliographystyle{plain}
\providecommand{\href}[2]{#2}
%

\bibliography{membrane_bib}

\raggedright

\end{document}